# Colloidal binary mixtures at fluid-fluid interfaces under steady shear: structural, dynamical and mechanical response

Ivo Buttinoni,*[a] Zachary A. Zell,[b], Todd M. Squires,[b] and Lucio Isa[a]

We experimentally study the link between structure, dynamics and mechanical response of two- dimensional (2D) binary mixtures of colloidal microparticles spread at water/oil interfaces. The particles are driven into steady shear by a microdisk forced to rotate at a controlled angular velocity. The flow causes particles to layer into alternating concentric rings of small and big colloids. The formation of such layers is linked to the local, position-dependent shear rate, which triggers two distinct dynamical regimes: particles either move continuously ("Flowing") close to the microdisk, or exhibit intermittent "Hopping" between local energy minima farther away. The shear-rate-dependent surface viscosity of the monolayers can be extracted from a local interfacial stress balance, giving "macroscopic" flow curves whose behavior corresponds to the distinct microscopic regimes of particle motion. Hopping Regions reveal a higher resistance to flow compared to the Flowing Regions, where spatial organization into layers reduces dissipation.

## 1 Introduction

Colloidal suspensions offer the unique opportunity to directly visualize structural response of a material to external perturbations at the level of individual constituents. In the specific case of colloidal systems subjected to shear flows, the mechanical response is very often coupled to the structure of the flowing suspension[1]. Local structural rearrangements may lead to drastic changes in macroscopic properties such as viscosity or elasticity[2]. Notable examples include shear-banding in colloidal crystals[3] and glasses[4], wall slip[5], the formation of hydroclusters in shear-thickening fluids[6] and alignment-layering transitions in shear-thinning fluids[7,8].

The simultaneous application of controlled stresses and the visualization of evolving morphologies in bulk materials typically requires confocal microscopes coupled to customized shear cells[9]. The necessity to scan large volumes across the geometry gap limits the range of accessible shear rates or restricts the observation to slices of material in proximity of solid boundaries[7,10]. These limitations can be circumvented by moving from bulk to truly two-dimensional (2D) systems. Particle monolayers can be produced by spreading colloids at macroscopically flat fluid interfaces[11], where interfacial forces trap microparticles irreversibly in the plane of the interface, and a range of attractive and repulsive interactions can be harnessed to control the interface microstructure[12]. For example, dipolar electrostatic repulsion, induced by the inhomogeneous distribution of charges across a water/oil interface[13], drives the formation of loosely-packed crystalline or glassy monolayers, with inter-particle distances reaching several particle diameters. More specifically, crystals are usually obtained when the colloids are monodisperse[14] or when the system is driven towards equilibrium[15,16], whereas polydisperse suspensions typically form glassy assemblies [17,18].

Experiments studying extensional[19] and steady shear flows[20,21] of interfacial monolayers showed that these 2D-colloidal crystals can be distorted by subjecting the interface to mechanical stresses. By analogy with shear experiments in bulk, deformations stem from local, cooperative, rearrangements which can induce the monolayers to align along slip planes[20]. In spite of the considerable importance of these discoveries and their strong applied implications in the engineering of particle-stabilized emulsions and foams[11], the experimental study of 2D-shear-induced structuring has so far been limited to monodisperse systems.

A first question that arises is the following: How do 2D binary suspensions restructure in the presence of steady shear flows? A second set of new questions addresses the interplay between these structures, the dynamics of single particles inside the potential landscape and the overall mechanical response to shear. How does the motion of individual particles vary when alignment-layering transitions occur? In which way are the global interface structure and the individual particle motion related to the local mechanical behavior of the 2D suspension?

In this article, we study loosely-packed binary (i.e., different par-ticle sizes) monolayers of micron-sized polystyrene (PS) spheres at various area fractions under continuous shear, applied via a microdisk rotating over a broad angular frequency range. Monolayers are prepared at water/decane interfaces where, in the absence of shear, the large particles assemble into ordered lattices while the overall structure (big and small particles altogether) does not show any long-range order. We demonstrate that the monolayers respond to shear by separating into series of alternating rings of large and small particles around the disk. By looking at the single-particle motion of the large beads we also find that the shear-induced structure is tightly coupled to both the dynamical and the mechanical response of the complex interface. Ordering under flow reduces the local viscosity, so that the interface behaves as a 2D shear-thinning fluid in a region close to the disk, similarly to what has been reported for bulk colloidal systems [8]. Beyond this layering region, the material adopts another flow modality whereby the strain propagates in a series of "Hopping" events between local energy minima [22–24]. The motion of the monolayers becomes hereby defect-mediated in analogy, for instance, to frictional motion across ordered substrates [25].

[a] *Laboratory for Interfaces, Soft matter and Assembly, Department of Materials, ETH Zurich, Switzerland. Tel: +41 44 632 63 89; E-mail: ivo.buttinoni@mat.ethz.ch.*
[b] *Department of Chemical Engineering, University of California, Santa Barbara, Santa Barbara, California 93106-5080, United States.*

The remainder of the paper is organized as follows: in Section 2 we present details of our interfacial colloidal system and the magnetic setup used to apply continuous shear to the interface. In Section 3 we report experimental results on the structural, the dynamical and the mechanical response of colloidal monolayers to steady shear and we emphasize the close connection between them. In Section 4 we discuss the results and the validity of our theoretical minimal model by addressing the role of the subphase. Finally, in Section 5 we summarize with our conclusions.

## 2 Experimental methods

Bidisperse colloidal monolayers are prepared by spreading sulfate PS particles at a flat water/decane interface. Interfacial shear is established by rotating circular magnetic probes at different frequencies. In this section we detail the experimental procedure.

Experiments are carried out in a custom-built cell, sketched in Fig. 1(a). A milliliter droplet of water is added at the bottom of the sample cell and its edge is strongly pinned at the rim of an aluminum funnel (aperture diameter 0.5 cm). An individual magnetic probe is later picked and deposited at the water/air interface using a sharp glass tip. We employ 'Janus microbuttons' (radius $R = 50$ $\mu$m, thickness 2 $\mu$m) as magnetic probes, which are fabricated from SU-8 photoresist by photolithography [26]. On top of the photoresist we sputter 200 nm of nickel followed by 10 nm of gold. The former renders the microdisk ferromagnetic, whereas the latter allows facile hydrophobic functionalization of the top side using a fluorothiol solution. Once the disk is inserted at the water/air interface we carefully pour n-decane on top to create an oil/water interface. The depth of the 2 phases is roughly 0.5 cm and a slight downward curvature of the interface is maintained so that the magnetic probe sits at the center of the cell by gravity. The curvature is later removed by adding a small amount of water to the sub-phase to ensure that the shear experiments are conducted at a macroscopically flat interface. Surfactant-free sulfate PS-particles (Interfacial Dynamics, USA, diameters $d_B = 4$ $\mu$m and $d_S = 1$ $\mu$m, number ratio 1:2) are spread directly at the

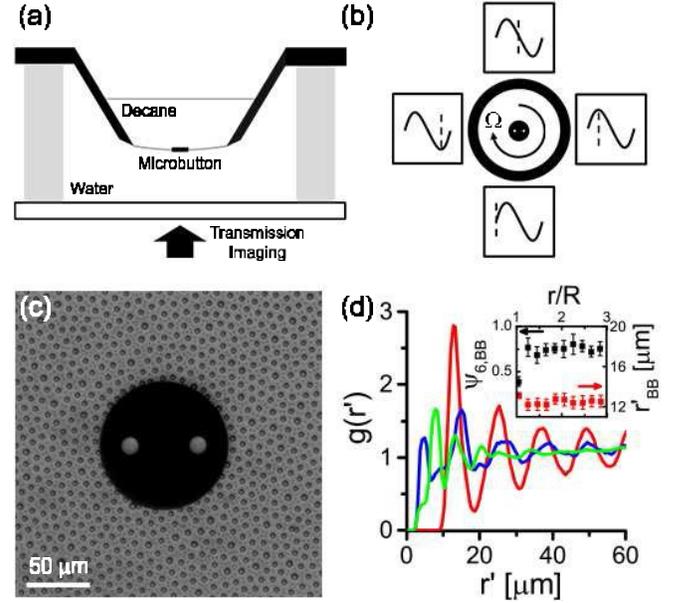

**Fig. 1** (a) Side view of the schematics of the experimental cell. The microprobe is centered at the w/o interface and colloids (not shown) are then spread at the interface with a micropipette. (b) Top view of the schematics of the setup. Two pairs of electromagnets create the magnetic field needed to rotate the microdisk. (c) Close-up snapshot of a bidisperse monolayer ($d_B = 4$ $\mu$m and $d_S = 1$ $\mu$m, $\varphi = 0.14$) around a magnetic probe ($R = 50$ $\mu$m) before shear is applied. (d) Corresponding pair correlation functions $g(r^t)$ plotted in units of the relative distance $r^t$ between big particles only (red), small particles only (blue) and big and small particles (green). Inset: order parameter $\psi_{6,BB}$ (black) and average inter-particle distance $r^t_{BB}$ (red) of big particles as a function of the normalized distance $r/R$ from the disk center.

water/decane interface using a 60:40 water:isopropanol mixture and a precision micropipette. Data are presented for experimental area fractions $\varphi$ (defined for convenience as the area occupied by the large particles) ranging between 0.04 and 0.20.

The magnetic setup schematically shown in Fig. 1(b) consists of four electromagnets controlled by two independent amplifiers [27,28]. Steady rotation of the magnetic microdisk, at frequencies ($\Omega/2\pi$) between 0.1 and 8 Hz, is achieved by applying a 90 degrees-phase delay between neighboring coils. The frequency of the driving current corresponds to the rotational frequency of the magnetic microdisk (extracted tracking the position of the holes), as verified by an initial calibration. The interface is imaged in transmission using 10× and 20× long-working distance objectives and snapshots are recorded with a CCD camera at 60 frames/s. The recorded image sequences are finally analyzed using custom Matlab codes in order to extract the positions of the particles in each frame.

## 3 Results

### 3.1 Structure of quiescent monolayers

After spreading, the colloidal particles self-assemble (at all area fractions $\varphi$ reported here) into non-closed packed structures due to electrostatic repulsion. Fig. 1(c) shows a typical monolayer at $\varphi = 0.14$ in the proximity of a magnetic probe. We always observe a layer of particles attached irreversibly to the disk; this corona facilitates no-slip boundary conditions at the probe's edge, which are an important prerequisite for velocity profile measurements.

The circular geometry of the disk distorts the monolayer structure only very close to the disk, corresponding to the first 1-2 layers located at few microns from the edge of the probe, due to the long-range softness of the inter-particle interactions. The effect of this geometrical perturbation on local ordering is illustrated in the inset of Fig. 1(d), where the hexagonal order parameter $\psi_{6,BB}$ and the average inter-particle distance $r^t_B$ of big particles are plotted as a function of the radial distance $r$ from the disk center. Significant deviations appear only in the immediate proximity of the microprobe (first data point for both curves). Hence, even though the shape of the probe is incommensurate with any local crystalline arrangement of the big particles, the disk does not perturb the interface microstructure beyond 1-2 lattice constants. Throughout this work, the microstructural rearrangements and surface velocity fields are measured in regions that are significantly larger than these one or two layers.

Outside of the perturbed region immediately adjacent to the microdisk, colloids self-assemble into binary structures where global crystallization is suppressed by the presence of small particles[17,29,30] even though large particles still maintain some long-range hexagonal order. The pair correlation functions $g(r^t)$ calculated from the particle positions (Fig. 1(d)) confirm that a series of peaks at well-defined inter-particle distances occurs when considering the large particles only (red curve). Instead, long-range order is lost when $g(r^t)$ is computed for other combinations, i.e., small and big particles (green) and small particles only (blue).

### 3.2 Structural response

The macroscopic interface structure displays a drastic change when rotational shear is applied compared to the quiescent case. Fig. 2 shows shear-induced structuring and the corresponding flow profiles in a binary monolayer ($\varphi = 0.14$) sheared at different angular frequencies $\Omega/2\pi$. The applied shear causes the formation of concentric layers, i.e., particles re-order and form concentric rings around the magnetic probe. Such layering is evident in the probability distributions of particle radial positions from the center of the rotating probe $P(r/R)$ in Figs. 2(b) and (c), in which each peak marks the position of a layer. In particular, Fig. 2(b) shows the radial position of the large particles both before and some time after 1.5 Hz probe rotation starts. At t = 0 s, just before shear starts, the monolayer is homogeneous over the entire interface, with the exception of the first 1-2 layers around the disk, which locally deforms the structure as previously described. After five seconds of disk rotation, multiple peaks (rings) have formed, extending significantly further away from the disk, indicating that the particles rapidly align with the external flow in the regions where the shear is sufficiently large to cause structural rearrangements. Further away from the disk the layering is lost. No further change in the structure of the monolayer is seen when analyzing data at longer times (e.g., at t = 10 s), indicating that steady state is rapidly reached within a few seconds.

Increasing the rotation rate $\Omega$ causes a greater number of rings to form (red and blue data in Fig. 2(c) correspond to 0.3 and 1

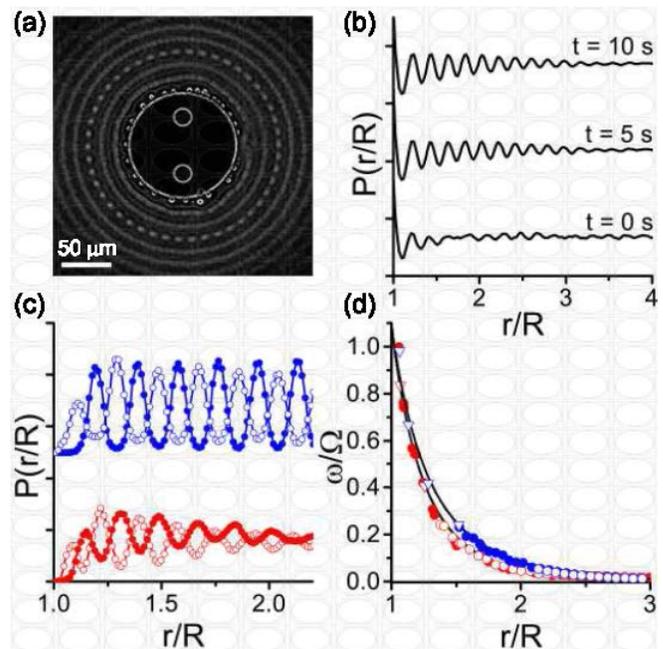

**Fig. 2** (a) Long-exposure image of a binary monolayer ($\varphi = 0.14$) sheared at 3 Hz in the co-moving reference of the rotating disk, highlighting the flow-induced structuring around the disk. (b) Normalized radial position probability distributions $P(r/R)$ of large particles at 0, 5 and 10 seconds after the disk starts to rotate at 1.5 Hz. (c) Steady state $P(r/R)$ for a monolayer sheared at 0.3 Hz (red data) and 1 Hz (blue data). Empty and filled symbols denote 1 $\mu$m and 4 $\mu$m particles, respectively. (d) Normalized angular velocity profiles corresponding to (c), obtained by tracking large and small particles (filled and open circles) and by image correlation (empty triangles), as discussed in the ESI.

Hz microdisk rotations within the same monolayer). Probability distributions curves are hereby extracted from both the positions of large (filled symbols) and small (empty symbols) particles. Remarkably, binary monolayers respond to the applied shear by separating into alternating layers of small and big colloids.

### 3.3 Dynamical response

Measured angular velocity profiles $\omega(r)$, normalized by the disk rotation rate $\Omega$, are shown in Fig. 2(d) for the 0.3 and 1 Hz rotations shown in 2(c). Two distinct methods were used to measure $\omega(r)$: direct tracking of large and small particles (filled and open circles) and image correlation (empty triangles), both of which give consistent results. In the first case, the angular velocity profiles are obtained by calculating the angular displacement of each particle within two consecutive frames and by averaging among the particles located at the same distance $r/R$ from the disk center. When image correlation methods are used, the local $\omega(r)/\Omega$ is calculated by finding the angle that maximizes the correlation between two circular stripes of the image centered around $r$ (see ESI for additional details). Correlation methods are required at high frequencies, when the standard tracking algorithms fail due to the fact that particle displacements between two consecutive frames become too large. Notably, the big and small particles follow the same velocity profiles. In what follows, we track the big particles alone, enabling lower-magnification objectives to be employed and

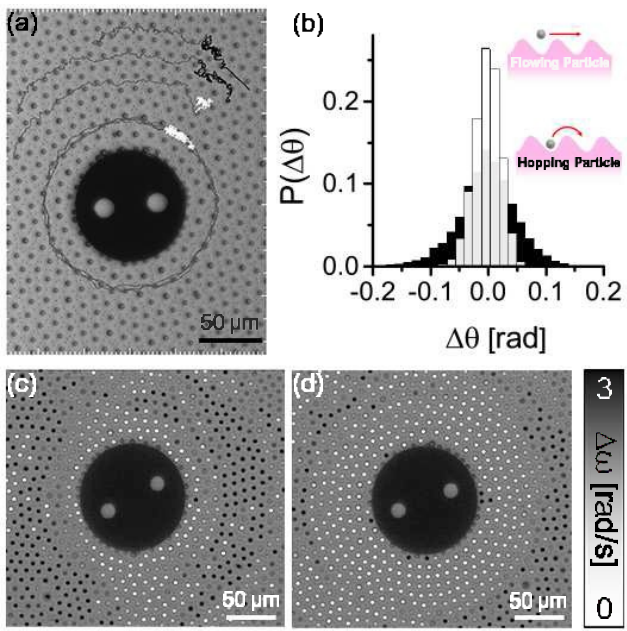
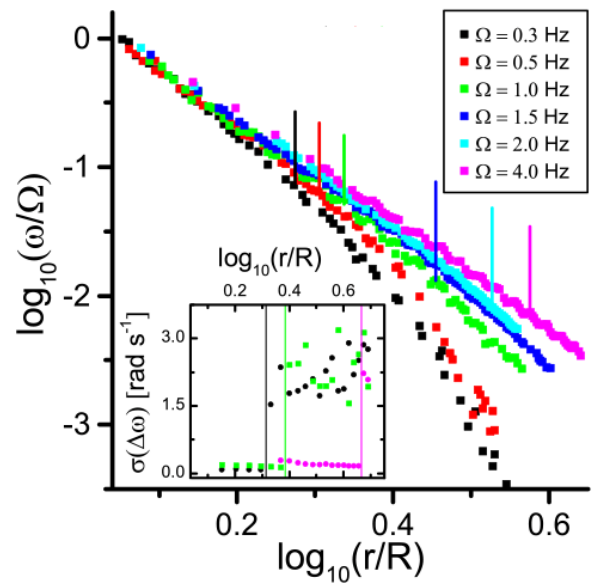

**Fig. 3** (a) Examples of trajectories in the lab (dark grey lines) and in the co-moving reference frame of the flow (black HR and white FR) for a monolayer sheared at 0.3 Hz. (b) Probability distributions of the angular displacements in the co-moving reference frame for two particles with initial radial positions $r/R = 1.3$ (white, FR) and 2.4 (black, HR). $\Omega/2\pi$ = 1.2 Hz and the displacements are measured over 700 frames. (c-d) $\Delta\omega$ (angular velocity in the co-moving frame) in the monolayer for two snapshots of increasing $\Omega$: (c) $\Omega/2\pi$ = 0.1 Hz and (d) $\Omega/2\pi$ = 0.3 Hz. Particles are colored according to their individual values of $\Delta\omega$ relative to the grey-scale on the right. Increasing $\Omega$, more particles enter the FR.

**Fig. 4** Log-log plot of the normalized angular velocities vs. $r/R$ for a binary monolayer with $\varphi = 0.14$ sheared at different probe rotations $\Omega$. Inset: normalized standard deviation of the angular velocity distributions in the co-moving reference frame as a function of $r/R$ (for clarity, only three frequencies are shown). The solid vertical lines mark the FR to HR transition.

broader areas of the monolayer to be imaged and tracked. As discussed in the next section, the presence of small particles does not affect the rheological properties of the monolayer (velocity profiles and flow curves for monodisperse systems are in Fig. 3 of the ESI).

Qualitatively distinct behaviors can be identified by comparing particles initially located at different distances from the disk (Fig. 3(a)) in the laboratory and co-rotating reference frames. The latter is obtained by subtracting the average angular motion at a given $r/R$ from the particle coordinates at the same radial position. White trajectories denote particles in a "Flowing" regime (FR), where the concentric rings are formed and the particles are advected by the flow. These particles are unlocked from their potential minima and move freely within the energy landscape[22]. Because FR colloids move with the same average speed as the surrounding shear flow, their trajectories resemble random walks when evaluated in the co-moving frame. Particles located farther from the disk move instead in a markedly different way. The black trajectories show that these particles do not flow smoothly, but rather hop occasionally in different directions. Trajectories in the co-moving frame for particles in this "Hopping" regime (HR) are thus no longer simple random walks; HR particles are trapped within a local potential minimum for a time, occasionally hopping into a neighboring minimum, reminiscent of zig-zag displacements[22] or crystallite rotations[20] in planar shear of monodisperse systems. Such intermittent hopping gives rise to angular displacements $\Delta\theta$ in the co-moving reference frame that are distributed much more broadly in the HR than in the FR (Fig. 3(b)).

This analysis can be performed for each particle in the monolayer. Figs. 3(c-d) show the relative angular speeds $\Delta\omega_i$ of each particle $i$, measured relative to the average speed $\langle\omega(r/R)\rangle$ at a radial distance $r/R$, defined as $\Delta\omega_i(r/R) = \omega_i(r/R) - \langle\omega(r/R)\rangle$. FR particles travel at the average flow speed, so that $\Delta\omega \approx 0$, and appear white. HR particles, on the other hand, travel with speeds that differ significantly from $\langle\omega\rangle$, and appear dark. The FR-HR transition is not smooth, as evidenced by abrupt increases in the standard deviation of $\Delta\omega$ normalized by the number of particles at $r/R$ (inset to Fig. 4). The radial location of the FR-HR transition (vertical lines in Fig. 4) depends on the interfacial shear stress imposed by the rotating disk, increasing with $\Omega$.

Broad features of these angular velocity profiles correlate directly with qualitative changes in the flow behavior. Fig. 4 shows the radial decay of the angular speed $\omega$ (normalized by $\Omega$) of a $\varphi = 0.14$ monolayer at different probe rotations $\Omega$ (analogous results for other $\varphi$ can be found in the ESI). In all cases, the interfacial velocity profile shows a no-slip coupling with the rotating disk ($\omega(R) = \Omega$). At large $\Omega$ (e.g., pink curve, $\Omega/2\pi = 4$ Hz), the monolayer is in the FR in almost the entire field of view, and $\omega$ shows a simple power-law decay. At smaller $\Omega$ (e.g., black data, $\Omega/2\pi = 0.3$ Hz), two distinct decays appear: a first power-law region close to the disk and a second, steeper decay at larger distances. Remarkably, the radial distance for the transition between the two slopes corresponds directly to the location of the FR-HR transition defined by the jump of $\sigma(\Delta\omega)$ (i.e., vertical lines in Fig. 4). This abrupt steepening of the decay in $\omega(r/R)$ reveals an increased resistance to deformation and flow that occurs when going from the

FR to the HR.

## 3.4 Mechanical response

Because the velocity field within the (2D) monolayer is not homogeneous, one can not determine the surface shear viscosity by simply dividing shear stress by shear rate. Indeed, the flow around the microdisk is effectively a Couette rheometer with an infinite gap. Because the velocity profile is measured directly, however, the surface shear viscosity can be determined so long as certain assumptions hold (which must be checked *a posteriori*). In particular, if the local flow is interfacially-dominated, then the surface shear viscosity $\eta_s$ can be determined from the local shear rate $\dot{\gamma}(r)$ according to

$$\eta_s(\dot{\gamma}) = \frac{\sigma_s(r)}{\dot{\gamma}(r)}, \quad (1)$$

where $\sigma_s(r)$ is the local surface shear stress on the monolayer[31].

While the surface shear rate $\dot{\gamma}(r)$ is straightforward to measure from measured velocity fields, the surface shear stress is not. If the Boussinesq number $Bo$ is large, however, shearing the monolayer requires significantly stronger stresses than shearing the subphase. $Bo$ is in fact defined as $Bo = \eta_s/(\eta_b R)$ and describes the importance of sub-phase contributions to the shear of complex interfaces[28,32], with $\eta_s$ and $\eta_b$ the surface and bulk viscosities, respectively. Therefore, in the high $Bo$ limit, the stress decay can be determined from a simple (2D) stress balance. In a given experiment, a torque

$$\tau = 2\pi R^2 \sigma_0 \quad (2)$$

is applied to the microbutton, which is transmitted to the monolayer in the form of a surface shear stress $\sigma_0$, exerted along the disk perimeter (with length $2\pi R$) and with a lever arm $R$. Assuming the surface shear stress to dominate over the subphase stresses, the interfacial torque is conserved for radii $r > R$. This implies that the surface shear stress in any experiment decays like

$$\sigma_{r\theta} = \sigma_0 \frac{R^2}{r^2}. \quad (3)$$

A more formal derivation of this relation follows from the momentum equation on the surface, which holds (within the continuum approximation)

$$\hat{\theta} \cdot (\nabla \cdot \sigma_{r\theta}) + f_\theta(r, u_B, u_\theta) = 0, \quad (4)$$

where $f_\theta$ is the viscous stress exerted on the monolayer by the subphase flow. Here we have assumed the surface stress to be given uniquely by the tangential component $\sigma_{r\theta}$, since all flow is azimuthal, and depends only on $r$. In cases where the surface shear stress significantly exceeds the subphase drag $f_\theta$, Eq. (4) reduces to

$$\frac{\partial \sigma_{r\theta}}{\partial r} + 2\frac{\sigma_{r\theta}}{r} \approx 0, \quad (5)$$

which is solved by

$$\sigma_{r\theta} = \frac{\sigma_0 R^2}{r^2}, \quad (6)$$

where $\sigma_0$ is the surface shear stress at the disk boundary.

Under the interfacially-dominated assumption, then, the surface shear stress in any monolayer is known up to the multiplicative constant $\sigma_0$. Measuring the azimuthal velocity field $u_\theta(r)$, or equivalently the angular velocity

$$\omega(r) = \frac{u_\theta}{r}, \quad (7)$$

allows the surface shear rate

$$\dot{\gamma}(r) = r\frac{\partial}{\partial r}\left(\frac{u_\theta}{r}\right) \quad (8)$$

to be extracted from measured velocity profiles for all $r$.

Once $\sigma_{r\theta}$ and $\dot{\gamma}$ have been measured, the local surface shear viscosity

$$\eta_s(\dot{\gamma}) = \frac{\sigma_{r\theta}(r)}{\dot{\gamma}(r)} = \frac{\sigma_0 R^2}{r^2 \dot{\gamma}(r)}. \quad (9)$$

can be extracted as a function of $r$ (and therefore $\dot{\gamma}$).

Fig. 5(a) shows the flow curves obtained from the velocity profiles in Fig. 4 following the method detailed above. Like the surface shear stress $\sigma_{r\theta}$, the surface shear viscosity is known up to a single multiplicative constant $\sigma_0$. While the surface shear stress $\sigma_0$ at the microdisk boundary ($r = R$) changes with $\Omega$, it is constant for each rotational frequency $\Omega$. Each rotation frequency $\Omega$ thus establishes approximately 30 distinct values of $\dot{\gamma}$, and therefore $\sim$ 30 distinct *local* measurements of $\eta_s(\dot{\gamma})$. The range of shear rates driven at one $\Omega$ overlaps substantially with range of shear rates driven at the next $\Omega$, whereas only one "fitting" parameter $\sigma_0$ can be chosen to shift the data. For each $\Omega$, then, a value of $\sigma_0$ is chosen to maximize the overlap of the measured $\eta_s(\dot{\gamma})$ curves with the rest of the $\Omega$ measurements. This way flow curves from different frequencies can be superimposed to form a master curve (Fig. 5(a), inset) valid for any $\Omega$. If this approach works, and the surface shear viscosity $\eta_s(\dot{\gamma})$ is indeed an intrinsic material property of the monolayer, then one expects to measure a single, master curve $\eta_s$ vs. $\dot{\gamma}$ for all experiments. As seen from Fig. 5(a), the flow curves extracted in this way do indeed collapse onto individual master curves, supporting the approach.

In the FR, measured velocity profiles exhibit a simple power-law decay,

$$\omega(r) = \frac{u_\theta}{r} = \Omega \left(\frac{r}{R}\right)^{-N} \quad (10)$$

so that

$$\dot{\gamma}(r) = -N\Omega \left(\frac{r}{R}\right)^{-N} = -N\omega(r) \quad (11)$$

In this case, using the constitutive relation $\sigma_{r\theta} = \eta_s(\dot{\gamma})\dot{\gamma}$ and extracting $r$ from Eq. (11), the surface viscosity also takes a power-law form

$$\eta_s(\dot{\gamma}) = \frac{\sigma_0}{(\Omega N)^{2/N}} \cdot \dot{\gamma}^{\left(\frac{2}{N}-1\right)}. \quad (12)$$

Distinct shear rate decays are measured for the FR and HR sections of the monolayer giving distinct decays for $\eta_s$. The arrows reported in the inset of Fig. 5(a) mark the shear rates calculated at the FR-HR transition found previously (Fig. 4). Notably, the critical shear rate at the FR-HR transition matches reasonably well for all probe rotation rates, confirming that this transition reflects an intrinsic material property of the monolayer. We emphasize

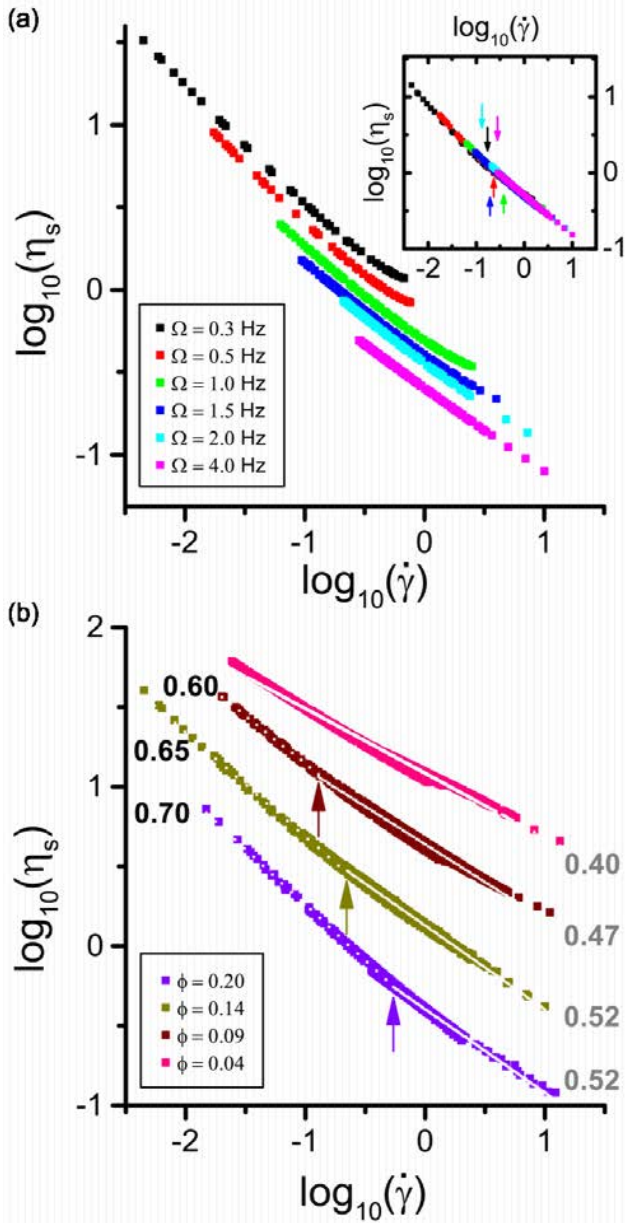

**Fig. 5** (a) Log-log plot of the surface viscosity corresponding to the profiles in Fig. 4 (colors are chosen accordingly) as a function of the local shear rate. Inset: Master flow curves obtained shifting all the curves on top of each other. (b) Master flow curves calculated for all densities and shifted for clarity. The lines and the numbers show power-law fits and (absolute) exponents in the FR (solid lines, grey numbers) and in the HR (dotted lines, black numbers). Arrows denote the average position of the FR-HR transition.

here that these flow curves are obtained locally, following an approach inspired by Goyon *et al.*[31], where by combining the overall stress balance at the interface and the local, measured shear rate, we measure different stress-strain relations, and thus viscosities, simultaneously and at different positions in the sample. The measurements in our "non-rheometric" infinite-gap Couette rheometer are uniquely enabled by the fact that we measure the local flow field, and couple it to the stress profile.

Analogous flow curves measured for different monolayer packings $\varphi$ (Fig. 5(b)) reveal the FR-HR (arrows) transition to occur at higher critical shear rates as $\varphi$ increases, as expected. In all cases we observe that the monolayers exhibit shear thinning, with different exponents, and thus material response, in the FR and HR regions. Fig. 5(b) shows explicitly power-law fits to the shear-thinning surface viscosity for FR (white solid lines) and HR (white dotted lines) portions and the (absolute) slopes of the fitting are reported next to the curves in black and grey, respectively. The coupling between structuring under flow and mechanical response leads to a lesser resistance to flow in the FR, where the concentric particle layers are found.

Finally, an experimental test has been performed using a monodisperse monolayer made solely of large particles at $\varphi = 0.19$ (data in the ESI). Comparison with bidisperse data at similar surface concentration (purple curve in Fig. 5(b), $\varphi = 0.20$) show no significant quantitative differences, thus strongly suggesting that the large particles bear most of the stress in the monolayers[33].

## 4 Discussion

The appearance of layers as a result of shear-induced rearrangements shown in Fig. 2 has been reported in several shear-thinning fluids[7,8,20]. In particular, in colloidal monodisperse suspensions, particles organize into layers in order to flow with less resistance. As opposed to monodisperse suspensions, experimental work addressing the layer formation in binary mixtures under shear is significantly lagging behind. Nonetheless, numerical simulations done by Löwen *et al.*[34] have envisaged that 2D-binary suspensions driven by external fields, including shear flows[35], may arrange into lanes of the same type of particles moving collectively with the field. In this way, and in the absence of vertical motion as in the case of particles trapped at fluid-fluid interfaces, the suspension maximizes transport parallel to the flow[34]. Our experimental findings confirm that there is a coupling between structure and flow, where the two conspire to reduce viscosity. In particular, in our experiments, the coupling happens locally and not on a global scale. We emphasize therefore here that our results are distinctively different to the case of standard shear-banding materials. In the latter case, the material develops bands of different viscosity in response to uniform shear[36], while in our case, given the geometry of the rotating probe, particles within the monolayer at different distances from the disk edge experience different shear stresses, and thus exhibit a different local rheological response.

The rheological information presented above requires nonetheless some care and implies subtleties in its interpretation. The viscosity curves presented in Fig. 5 have been computed assuming the shear stress transmitted by the disk to be borne entirely within the monolayer. However, subphase contributions might be present and influence the results. Since the measurements shown in Fig. 5 provide access to the surface viscosity only up to a multiplicative constant (the stress scale $\sigma_0$), hydrodynamic arguments must be used to check the validity of the above-mentioned assumption. Velocity profiles within the interface plane around rotating disks for Newtonian interfaces with negligible surface viscosities, *i.e.*, in the subphase-dominated limit ($Bo \ll 1$), have been calculated[37] showing a $u_\theta \sim r^{-2}$, or $\omega \sim r^{-3}$ decay. Conversely, the velocity profiles of a Newtonian interface with high surface viscosity ($Bo \gg 1$) give a $u_\theta \sim r^{-1}$, or $\omega \sim r^{-2}$ decays[28]. As previously mentioned (Eq. (10) and Fig. 4) measured velocity profiles in the FR reveal indeed power-law decays,

$$\omega(r) \sim r^{-N} \quad (13)$$

and therefore examining the values of the exponents $N$ yields significant information on rheology of the monolayers. Any angular velocity field at the interface that that decays more rapidly than $r^{-3}$ cannot arise from a subphase-dominated flow alone, but directly implies the existence of non-negligible surface viscosities. Additionally, any values of $N > 3$ also implies a shear-thinning interface, where the viscosity decreases faster and where the shear rates are higher compared to the Newtonian case. In the specific case of a 2D shear-thinning suspension, $N > 3$ therefore also unambiguously reflects rheological response due to stresses within the monolayer. The green data ($\varphi = 0.14$) of the graph in Fig. 6 illustrates the $\Omega$-dependence of the power-law exponents $N$ of the velocity profiles presented in Fig. 4. The plot shows that, at that particular packing fraction, $N$ is always significantly greater than 3 for all $\Omega$. Fig. 6 also includes the exponents $N$ measured using monolayers at different $\varphi$ (the corresponding velocity curves are reported in the ESI). We note that all monolayers with $\varphi > 0.1$ shear thin with $N > 4$ at all experimental $\Omega$, the monolayer with $\varphi = 0.09$ shows $N > 4$ for most $\Omega$, and even the system with $\varphi = 0.04$ yields an exponent $N$ appreciably above 3 at low frequencies. These results confirm the predominance of interfacial effects in most of our data.

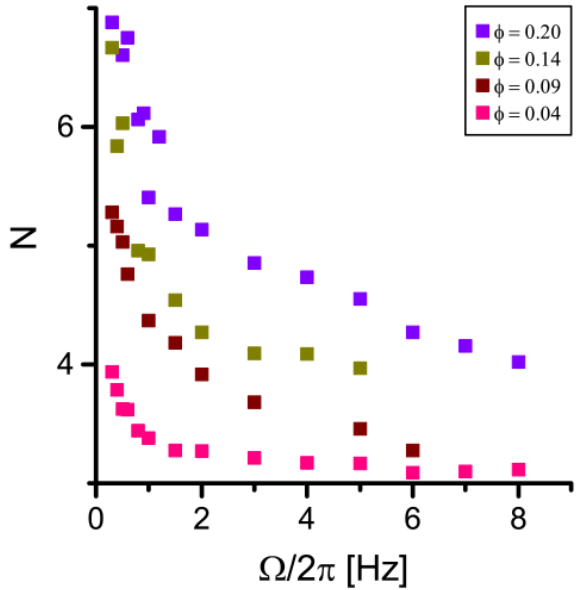

**Fig. 6** Exponents of the decay of the FR part of the angular velocity profiles $\omega(r)$ as a function of $\Omega$ for different packings $\varphi$ (colors are chosen to match with the data in Fig. 5(b)). The green data ($\varphi = 0.14$) correspond to the velocity profiles in Fig. 4.

Further justification for the interface-dominated stress profiles assumption leading to Eq. (5) is provided by the flow curves (Fig. 5(b)) themselves. In the $Bo \ll 1$ limit the interfacial shear rate $\dot\gamma$ would also decay like $r^{-3}$, which would correspond to an apparent surface viscosity, using Eq. (12),

$$\eta_s^{app} \sim \dot\gamma^{-1/3}. \quad (14)$$

In most cases, the measured surface shear viscosities in the FR shear-thin much more strongly than this, indicating unambiguously the presence of interfacial stresses. The interfacially-dominated assumption may break down at some of the highest shear rates and lowest concentrations, where $N \to 3$ and $\eta_s^{app} \sim \dot\gamma^{-1/3}$ (e.g., pink data in Figs. 6 and 5(b), at large $\dot\gamma$). For this set of data Eq. (12) leads to the appearance of an apparent surface viscosity due to the subphase stress contribution. The unavoidable presence of both surface and subphase stresses and their respective balance ultimately defines the system's rheological response. In particular, we expect that any suspension (dilute enough that it does not shear thicken), will show a Newtonian plateau with $\eta_s = const$ at sufficiently high rates[38]. In the case of our experiments at the lowest area fraction, the shear-thinning nature of the monolayer leads to a reduction of the interfacial viscosity such that the transition from interface to subphase-dominated flows ($Bo \sim 1$) happens before the high-shear-rate Newtonian plateau is reached. Hence the steady apparent decay of the interfacial viscosity at all rates. This case is in contrast to the case of some surfactant monolayers that exhibit Newtonian viscosities well in the $Bo \gg 1$ regime[28]. Nevertheless, it is worth noting that, while subphase contributions might affect the FR-slope of the viscosity curves in the extreme cases mentioned above, the shear rates at which the FR-HR transition is observed for all our other data have much smaller values and the FR branches present significant deviations from the -1/3 decay.

We can finally safely say that the presence of the probe does not affect the structure and the mechanical response of the monolayer. As we have described in section 3.1 the circular geometry of the disk induces very local deformations in the structure of the monolayer. In the first 1-2 layers from the disk edge the colloids position themselves at preferred positions, perturbing locally the lattice, even in the absence of flow. As shown in Fig. 1(c,d) and discussed in the corresponding section, this "splay" is very circumscribed and is overcome by the shear-induced structures already at small rotation frequencies. An additional proof that the probe does not affect significantly the monolayer response has been obtained by looking at the flow field generated by both circular, hexagonal and square probes (data shown in the ESI). For the two latter cases, the shear flow leads to the formation of circular layers, identical to the ones shown in Fig. 2(a) for the disk, after just a few lattice spacings away from the probe edge.

## 5 Conclusions

In conclusion, we have demonstrated that continuous, radially symmetric shear flow significantly restructures 2D binary colloidal monolayers, forming concentrically layered rings. Data extracted shearing binary mixtures absorbed at a liquid-liquid interface corroborate earlier numerical simulations predicting a shear-induced separation of the mixture into alternating layers of small and big particles[35]. The structural reorganization of the material directly corresponds to qualitative changes in the dynamical response of in-dividual particles, from Flowing to Hopping. In turn, this shift corresponds directly to a clear transition in the macroscopic mechanical properties of the surface. Surface shear viscosities extracted from interfacial velocity profiles measured at different probe rotations collapse onto $\phi$-dependent master flow curves, with critical shear rates for FR-HR transitions, consistent with intrinsic material properties. The structural, dynamical and rheological responses of these complex interfaces are clearly interrelated, highlighting the connection between morphological process and rheological behavior that must be considered when designing complex fluid interfaces. They furthermore reinforce the view that the macroscopic response of a material is intimately linked to the microscopic behavior of its constituents, a link that is particularly apparent for colloidal systems, where individual constituents can be directly followed.

An appealing outlook for our work addresses the response of such 2D systems to oscillatory perturbations. Previous work on the oscillatory rheology of colloidal monolayers has demonstrated that they behave as soft glassy materials[39]. Additionally, recent experiments performed by Keim *et al.*, using a needle interfacial shear rheometer combined to the visualization of the sheared material, made it possible to observe shear transformation zones appearing when a colloidal monolayer is subjected to a linear shear deformations[33,40]. We envisage the possibility to study the plastic/elastic response of colloidal monolayers under oscillatory shear applied by our magnetic microdisks while monitoring local rearrangements of the particles and thus shed additional light onto the mechanisms behind phenomena such the onset of yielding and plasticity in soft 2D materials.

## 6 Acknowledgements

The authors acknowledge Jan Vermant for inspiring dicussions, LI and IB acknowledge financial support from the SNSF grants PP00P2_144646/1 and IZK0Z2_142110/1, and TMS from the National Institutes of Health Grant HL-51177.

## References

1  J. Vermant and M. J. Solomon, *Journal of Physics: Condensed Matter*, 2005, **17**, R187.

2  D. T. Chen, Q. Wen, P. A. Janmey, J. C. Crocker and A. G. Yodh, *Condensed Matter Physics*, 2010, **1**, year.

3  I. Cohen, B. Davidovitch, A. B. Schofield, M. P. Brenner and D. A. Weitz, *Phys. Rev. Lett.*, 2006, **97**, 215502.

4  R. Besseling, L. Isa, P. Ballesta, G. Petekidis, M. E. Cates and W. C. K. Poon, *Phys. Rev. Lett.*, 2010, **105**, 268301.

5  P. Ballesta, R. Besseling, L. Isa, G. Petekidis and W. C. K. Poon, *Phys. Rev. Lett.*, 2008, **101**, 258301.

6  X. Cheng, J. H. McCoy, J. N. Israelachvili and I. Cohen, *Science*, 2011, **333**, 1276–1279.

7  M. Haw, W. Poon, P. Pusey, P. Hebraud and F. Lequeux, *Physical Review E*, 1998, **58**, 4673.

8  Y. Wu, D. Derks, A. van Blaaderen and A. Imhof, *Proc Natl Acad Sci U S A.*, 2009, **106**, 10564–10569.

9  L. Isa, R. Besseling, A. Schofield and W. Poon, *Advances in Polymer Science: High Solid Dispersions*, 2010, **236**, 163–202.

10  K. Jensen, D. A. Weitz and F. Spaepen, *Physical Review E*, 2014, **90**, 042305.

11  B. P. Binks and T. S. Horozov, *Colloidal Particles at Liquid Interfaces*, 2006.

12  F. Bresme and M. Oettel, *Journal of Physics: Condensed Matter*, 2007, **19**, 413101.

13  C. L. Wirth, E. M. Furst and J. Vermant, *Langmuir*, 2014, **30**, 2670–2675.

14  P. Pieranski, *Physical Review Letters*, 1980, **45**, 569.

15  A. D. Law, D. M. A. Buzza and T. S. Horozov, *Phys. Rev. Lett.*, 2011, **106**, 128302.

16  A. Law, T. Horozov and D. Buzza, *Soft Matter*, 2011, **7**, 8923–8931.

17  Ebert, F., Keim, P. and Maret, G., *Eur. Phys. J. E*, 2008, **26**, 161–168.

18  L. J. Bonales, F. Martinez-Pedrero, M. A. Rubio, R. G. Rubio and F. Ortega, *Langmuir*, 2012, **28**, 16555–16566.

19  E. J. Stancik, M. J. O. Widenbrant, A. T. Laschitsch, J. Vermant and G. G. Fuller, *Langmuir*, 2002, **18**, 4372–4375.

20  E. J. Stancik, G. T. Gavranovic, M. J. O. Widenbrant, A. T. Laschitsch, J. Vermant and G. G. Fuller, *Faraday Discuss.*, 2003, **123**, 145–156.

21  S. Barman and G. F. Christopher, *Langmuir*, 2014, **30**, 9752–9760.

22  W. Loose and B. J. Ackerson, *The Journal of Chemical Physics*, 1994, **101**, 7211–7220.

23  E. J. Stancik, A. L. Hawkinson, J. Vermant and G. G. Fuller, *Journal of Rheology (1978-present)*, 2004, **48**, 159–173.

24  E. Pratt and M. Dennin, *Physical Review E*, 2003, **67**, 051402.

25  T. Bohlein, J. Mikhael and C. Bechinger, *Nature materials*, 2012, **11**, 126–130.

26  S. Q. Choi, S. G. Jang, A. J. Pascall, M. D. Dimitriou, T. Kang, C. J. Hawker and T. M. Squires, *Advanced Materials*, 2011, **23**, 2348–2352.

27  S. Choi, S. Steltenkamp, J. Zasadzinski and T. Squires, *Nature communications*, 2011, **2**, 312.

28  Z. A. Zell, A. Nowbahar, V. Mansard, L. G. Leal, S. S. Deshmukh, J. M. Mecca, C. J. Tucker and T. M. Squires, *Proceedings of the National Academy of Sciences*, 2014, **111**, 3677–3682.

29 F. Ebert, G. Maret and P. Keim, *The European Physical Journal E: Soft Matter and Biological Physics*, 2009, **29**, 311–318.

30 R. Yamamoto and A. Onuki, *Physical Review E*, 1998, **58**, 3515.

31 J. Goyon, A. Colin, G. Ovarlez, A. Ajdari and L. Bocquet, *Nature*, 2008, **454**, 84–87.

32 G. G. Fuller and J. Vermant, *Annual review of chemical and biomolecular engineering*, 2012, **3**, 519–543.

33 N. C. Keim and P. E. Arratia, *Soft matter*, 2015, **11**, 1539–1546.

34 J. Dzubiella, G. Hoffmann and H. Löwen, *Physical Review E*, 2002, **65**, 021402.

35 H. Löwen, R. Messina, N. Hoffmann, C. N. Likos, C. Eisenmann, P. Keim, U. Gasser, G. Maret, R. Goldberg and T. Palberg, *Journal of Physics: Condensed Matter*, 2005, **17**, S3379.

36 S. M. Fielding, *Reports on Progress in Physics*, 2014, **77**, 102601.

37 F. C. Goodrich, *Proceedings of the Royal Society of London A: Mathematical, Physical and Engineering Sciences*, 1969, **310**, 359–372.

38 J. Mewis and N. J. Wagner, *Colloidal suspension rheology*, Cambridge University Press, 2012.

39 P. Cicuta, E. J. Stancik and G. G. Fuller, *Physical review letters*, 2003, **90**, 236101.

40 N. C. Keim and P. E. Arratia, *Physical review letters*, 2014, **112**, 028302.